\documentclass[12pt]{article}
\usepackage{amsmath,amsthm,amscd,amssymb}
\usepackage{latexsym}
\usepackage{epsf}
\usepackage{epsfig}
\begin{document}

\title{Nonlocal Lagrangians for Accelerated Systems}
\author{C. Chicone\\Department of Mathematics\\University of 
Missouri-Columbia\\Columbia, Missouri 65211, USA \and B. 
Mashhoon\thanks{Corresponding author. E-mail:
mashhoonb@missouri.edu (B. Mashhoon).} \\Department of Physics and
Astronomy\\University of Missouri-Columbia\\Columbia, Missouri 65211, USA}
\maketitle  

\begin{abstract}
Acceleration-induced nonlocality and the corresponding Lorentz-invariant nonlocal field equations of accelerated systems in Minkowski spacetime are discussed. Under physically reasonable conditions, the 
nonlocal equation of motion of the field can be derived from a variational principle of stationary action involving a nonlocal Lagrangian that is simply obtained by composing the local inertial Lagrangian with the nonlocal transformation of the field to the accelerated system.
The implications of this approach for the electromagnetic and Dirac fields
are briefly discussed.

\end{abstract}
\noindent \emph{PACS}: 03.30.+p; 11.10.Lm; 04.20.Cv\\
\emph {Keywords}: Relativity; Nonlocality; Accelerated observers 

\section{Introduction}
The basic nongravitational laws of physics are formulated with respect to ideal inertial observers; however, actual observers are accelerated. In particular, the observational basis of Lorentz invariance rests upon measurements performed by noninertial observers. It is therefore necessary to establish a connection between accelerated and inertial observers in Minkowski spacetime. 

In the standard theory of relativity, the extension of Lorentz invariance to noninertial observers rests upon the \emph{hypothesis of locality}, namely,
the assumption that an accelerated observer along its worldline is pointwise equivalent to an otherwise identical momentarily comoving inertial observer; hence, an accelerated observer can be replaced by a continuum of hypothetical momentarily comoving inertial observers. The physical origin and possible limitations of the hypothesis of locality have been discussed at length before (see, for example, \cite{L1} and the references therein).  If all physical phenomena could be reduced to pointlike coincidences of classical point particles and rays of radiation (``eikonal limit''), the locality hypothesis would be exact. 

To go beyond the hypothesis of locality, a nonlocal theory of accelerated observers in Minkowski spacetime has been developed that appears to be consistent with quantum theory~\cite{3,4,5,6}. Moreover, the consequences of this nonlocal theory are in agreement with observational data available at present. In particular, the nonlocal theory forbids the existence of a pure scalar (or pseudoscalar) radiation field~\cite{3,4,5}. 

Consider an accelerated observer in a global background inertial frame $\mathcal{S}$ and let $\psi$ be a fundamental field in spacetime. The noninertial observer along its worldline passes through a continuum of hypothetical comoving inertial observers. Let $\hat\psi(\tau)$ be the field measured by the comoving inertial observer at the event characterized by the proper time $\tau$. The ``local'' spacetime of such an inertial observer is related to the background inertial frame $\mathcal{S}$ by a Poincar\'e transformation $x^\alpha=L^\alpha_{\hspace{0.1in} \beta}\, x'^\beta+s^\alpha$; therefore, $\psi'(x')=\Lambda(L)\psi(x)$, where $\Lambda$ belongs to a representation of the Lorentz group. Thus, $\hat\psi(\tau)=\Lambda(\tau) \psi(\tau)$ along the worldline of the accelerated observer. 

Let $\hat\Psi(\tau)$ be the field that is actually measured by the accelerated observer at $\tau$. The hypothesis of locality postulates that $\hat\Psi(\tau)=\hat\psi(\tau)$. To construct a nonlocal theory, we note that the most general linear relation between  $\hat\Psi(\tau)$ and $\hat\psi(\tau)$, which is consistent with causality, is given by
\begin{equation}\label{eq:1}
\hat\Psi(\tau)=\hat\psi(\tau)+\int_{\tau_0}^\tau \hat k(\tau,\tau')\hat\psi(\tau')\,d\tau',
\end{equation}
where $\tau_0$ denotes the instant at which the observer's acceleration is turned on (see~\cite{3}). 

To avoid possible unphysical situations, we assume that the observer is accelerated only for a finite interval of proper time. Equation~\eqref{eq:1} involves spacetime scalars; thus, it is manifestly invariant under Poincar\'e transformations of the background spacetime. The kernel $\hat k$ is obtained from the acceleration of the observer.  It vanishes for an inertial observer and the nonlocal part of Eq.~\eqref{eq:1}  vanishes in the eikonal limit. Moreover,
the field seen by the accelerated observer involves a spacetime average over the past worldline of the observer, in agreement with the viewpoint developed by Bohr and Rosenfeld~\cite{4a}.

A simple consequence of Lorentz invariance for ideal inertial observers is that a basic radiation field can never stand completely still with respect to an observer (see~\cite{3}). This physical postulate---generalized to arbitrary accelerated observers---is used to determine the kernel in Eq.~\eqref{eq:1}, a process that is extensively discussed in~\cite{9,10,11,1,L11}. In the following, we simply assume that an appropriate kernel can be determined from the acceleration of the observer. 

Equation~\eqref{eq:1} is a Volterra integral equation of the second kind. A theorem of Volterra~\cite{12} states that if $\tau_1> \tau_0$ and the kernel is continuous in the proper time domain $\tau_0\le \tau'<\tau\le \tau_1$, then $\hat\psi$ is uniquely determined by $\hat\Psi$ in the space of continuous function on the interval $[\tau_0, \tau_1]$. A similar result was proved by Tricomi~\cite{13} in the Hilbert space of square-integrable functions. Thus, under mild assumptions, 
we have that $\hat\psi$ is uniquely determined by $\hat\Psi$; in fact, 
\begin{equation}\label{eq:2}
\hat\psi(\tau)= \hat\Psi(\tau)+\int_{\tau_0}^\tau \hat r(\tau,\tau')\hat\Psi(\tau')\,d\tau'
\end{equation}
for some new kernel $\hat{r}$ called the resolvent kernel (see~\cite{L14}). 

Let us assume that a field $\Psi$ exists such that for the accelerated observer under consideration here 
\begin{equation}\label{eq:3}
\hat\Psi(\tau)=\Lambda(\tau) \Psi(\tau).
\end{equation}
Substituting this relation into Eq.~\eqref{eq:2}, we find that
\begin{equation}\label{eq:4}
\psi(\tau)= \Psi(\tau)+\int_{\tau_0}^\tau  r(\tau,\tau')\Psi(\tau')\,d\tau',
\end{equation}
where $r$ is related to the resolvent kernel $\hat r$  via 
\begin{equation}\label{eq:5}
r(\tau,\tau')=\Lambda^{-1}(\tau)\hat r(\tau,\tau')\Lambda(\tau').
\end{equation}

Let us now consider a congruence of accelerated observers and assume that  Eq.~\eqref{eq:4} is extended to the whole congruence so that $\psi$ is related to a field $\Psi$ by a nonlocal relation involving a suitable kernel $K$ via the integral equation 
\begin{equation}\label{eq:6}
\psi(x)=\Psi(x)+\int K(x,x') \Psi(x')\, d^4 x'.
\end{equation} 
In view of the results of Volterra and Tricomi,  $\Psi$ is expected to be uniquely determined by $\psi$ under mild mathematical assumptions; and, the uniqueness would seem to be demanded on physical grounds. However, we note
that this uniqueness result is valid in a finite region of integration;
therefore, we imagine that all physical processes of interest take place in
a sufficiently large but finite spacetime domain outside of which the kernel $K$ vanishes.  The construction of such kernels is illustrated in section 6. Thus in what follows, we require that the kernel be supported in an open subset of $\mathcal{M} \times \mathcal{M}$ with compact closure, where $\mathcal{M}$ is the Minkowski spacetime. Under these conditions, we postulate that the kernel $K$ (which is determined by the acceleration of our congruence of observers)  is such that  $\Psi$ is uniquely determined by $\psi$. This postulate plays a fundamental role in the formulation of a variational principle for the field $\Psi$.

The local field $\psi$ satisfies a field equation that may be expressed as $\mathcal{O}_x[\psi](x)=0$, where $\mathcal{O}_x$ is a differential operator. By applying $\mathcal{O}_x$ to  Eq.~\eqref{eq:6}, it follows that $\Psi$ satisfies the equation  $\mathcal{O}_x[\mathcal{N}\Psi](x)=0$, where $\mathcal{N}$ is the nonlocal operator defined by
\begin{equation}\label{eq:7}
\psi=\mathcal{N}\Psi.
\end{equation}
Our postulate states that $\mathcal{N}$ is an invertible operator in the space of candidates for fields. 

Explicit nonlocal field equations have been obtained in some simple cases (see~\cite{1,L14,star}). The nonlocal Maxwell and Dirac equations are discussed in sections 4 and 5, respectively.  Also, we note an important feature of fields that satisfy these nonlocal field equations: nonlocality survives---in the form of the memory of past acceleration---even after the acceleration of an observer is turned off.

Suppose that the field equation  $\mathcal{O}_x[\psi](x)=0$ is obtained from the Euler-Lagrange equation
\begin{equation}\label{eq:8}
\frac{\partial \mathcal{L}}{\partial \psi}-\frac{\partial}{\partial x^\mu}\Big (
\frac{\partial \mathcal{L}}{\partial (\partial_\mu\psi)}\Big)=0
\end{equation}
associated with the variational principle of stationary action, $\delta A=0$, where
\begin{equation}\label{eq:9}
A[\psi] = \int \mathcal{L} (x,\psi,\partial_\mu\psi)\, d^4x.
\end{equation}
Our main purpose is to show that, under suitable conditions, the nonlocal field equation $\mathcal{O}_x[\mathcal{N}\Psi](x)=0$ can be derived by the principle of stationary action from the functional
\begin{equation}\label{eq:10}
\mathcal{A}[\Psi] = \int \mathcal{L} (x,\mathcal{N}\Psi(x),\partial_\mu\mathcal{N}\Psi(x))\, d^4x.
\end{equation} 
The main physical significance of this result is that it makes it possible to derive in a consistent manner the nonlocal field equations for \emph{interacting} fields; for instance, the interaction of charged Dirac particles with the electromagnetic field can be studied from the viewpoint of accelerated observers. 

To explain our method, we present a simple toy model in the next section followed by a general treatment in section 3. 

\section{Toy model}
To illustrate the variational analysis for the functional~\eqref{eq:10}, we consider a toy model consisting of a real massless scalar field $\phi(x)$ with the standard Lagrangian
\begin{equation}\label{eq:11}
\mathcal{L}=\frac{1}{2} \eta^{\mu\nu}\partial_\mu\phi\,\partial_\nu\phi,
\end{equation} 
where $\eta_{\mu\nu}=\mbox{diag}\, (1,-1,-1,-1)$ is the Minkowski metric tensor. Here $x$ stands for $x^\mu=(t,x^i)$ and units are chosen such that $c=\hbar=1$.
Let $\phi=\mathcal{N}\Phi$, where $\mathcal{N}$ is the nonlocal operator defined in display~\eqref{eq:7} and $\Phi$ is a real scalar field in our model. The field equation is  $\Box \phi=0$, where $\Box=\eta^{\mu\nu}\partial_\mu\partial_\nu$. We will show that the field equation for $\Phi$, $\Box (\mathcal{N}\Phi)=0$, is obtained by variation of the functional
\begin{equation}\label{eq:12}
\mathcal{A}[\Phi] = \int\frac{1}{2}\eta^{\mu\nu} \partial_\mu(\mathcal{N}\Phi)\partial_\nu(\mathcal{N}\Phi) \, d^4x.
\end{equation} 

Note that 
\begin{equation}\label{eq:13}
\delta \mathcal{A} = \mathcal{A}[\Phi+\delta\Phi]-\mathcal{A}[\Phi],
\end{equation} 
where $\delta\Phi$ is the variation of the field such that $\delta\Phi$ vanishes at the boundary of the spacetime region of interest. We have the equation
\begin{equation}\label{eq:14}
\delta \mathcal{A} =\int U\delta \Phi\, d^4x+\int \partial_\mu V^\mu \, d^4x,
\end{equation} 
where 
\begin{equation}\label{eq:15}
U=-\Box(\mathcal{N}\Phi)+ \int \eta^{\mu\nu}\frac{\partial\mathcal{N}\Phi(y)}{\partial y^\mu}  \frac{\partial K(y,x)}{\partial y^\nu} \, d^4y
\end{equation}
and 
\begin{equation}\label{eq:16}
V^{\mu}=\eta^{\mu\nu}\partial_\nu(\mathcal{N}\Phi)\delta\Phi.
\end{equation}
By Stokes' theorem, 
\begin{equation}\label{eq:17}
\int \partial_\mu S^\mu \, d^4x=\int  S^\mu d^3\Sigma_\mu, 
\end{equation}
where $d^3\Sigma_\mu=\frac{1}{3!}\epsilon_{\mu\alpha\beta\gamma}\, dx^\alpha\wedge dx^\beta\wedge dx^\gamma$, $\epsilon_{0123}=1$, and the latter integration in Eq.~\eqref{eq:17} is carried out over a closed three-dimensional boundary hypersurface in spacetime. On such a hypersurface, $\delta\Phi=0$ by definition; hence, the second integral in Eq.~\eqref{eq:14} vanishes.  Using the new field $G$ given by 
\begin{equation}\label{eq:18}
G(x)=-\Box[\mathcal{N}\Phi](x),  
\end{equation}
the formula for $U$ can be rewritten as 
\begin{equation}\label{eq:19}
U(x)=G(x)+\int G(y)K(y,x)d^4y+\int\frac{\partial}{\partial y^\mu}\big[\eta^{\mu\nu} \frac{\partial\mathcal{N}\Phi(y)}{\partial y^\nu} K(y,x) \big]\, d^4y.
\end{equation}

Because the kernel $K$ is nonzero only in a 
finite region of spacetime, the boundary hypersurface can be chosen such that the second integral in Eq.~\eqref{eq:19} vanishes via Stokes' theorem. Moreover, in the Hilbert space of square-integrable candidates for fields, $U$ can be written as 
\begin{equation}\label{eq:20}
U(x)=\mathcal{N}^\dagger G(x),
\end{equation}
where $\mathcal{N}^\dagger$ is the (Hilbert space) adjoint of $\mathcal{N}$. Thus, $\delta\mathcal{A}=0$ in Eq.~\eqref{eq:14} implies that $U=0$. By our postulate, $\mathcal{N}$ is invertible. It follows that $\mathcal{N}^\dagger$ is invertible (see~\cite{L15,L16}); therefore, $G=0$ and $\Box (\mathcal{N}\Phi)=0$ is the nonlocal equation of motion.

\section{Variational principle for accelerated\\ observers}
Our analysis of the variational principle for the toy model involving a real massless scalar field has a natural generalization to arbitrary Lagrangians that will be explained in this section. The fundamental result is that the nonlocal Lagrangian is simply the local Lagrangian composed with the nonlocal operator $\mathcal{N}$ that relates fields to those experienced by accelerated observers.

Consider a Hilbert space $\mathfrak{X}$ of square-integrable fields, each defined on some open bounded spacetime domain $\Omega$ with values in $\Omega\times \mathbb{R}^n$ with $n>1$. More generally, the fields may take values in the total space of a vector or tensor bundle; but, for simplicity, we will only consider trivial bundles here. Let $\mathfrak{M}$ denote the space of $n\times n$ matrices, let $\partial\Omega$ be the boundary of $\Omega$  and $K:\Omega\times\Omega\to \mathfrak{M}$ be defined so that the operator $\mathcal{N}:\mathfrak{X}\to \mathfrak{X}$ given by
\begin{equation}\label{eq:30}
(\mathcal{N}\Psi)(x):=\Psi(x)+\int_\Omega K(x,y)\Psi(y)\,d^4y
\end{equation}
is invertible and the kernel $K(x,y)$ is the zero matrix whenever $(x,y)$ is in the complement of $\Omega\times \Omega$.

Let $A:\mathfrak{X}\to \mathbb{R}$ be the functional 
\begin{equation}\label{eq:30.1}
A(\Psi)=\int_\Omega \mathcal{L}(x, \Psi(x),\nabla \Psi(x))\,d^4x
\end{equation}
associated with the local 
Lagrangian $\mathcal{L}:\Omega \times\mathbb{R}^n \times \mathbb{R}^{4 n} \to \mathbb{R}$ given by 
$(x,u,v)\mapsto \mathcal{\mathcal{L}}(x,u,v)$. 
The functional $A$ composed with $\mathcal{N}$ is the new functional
\begin{equation}\label{eq:31}
\mathcal{A}(\Psi):=(A\circ \mathcal{N})(\Psi)=\int_\Omega \mathcal{\mathcal{L}}(x,(\mathcal{N}\Psi)(x), \nabla(\mathcal{N}\Psi)(x))\,d^4x,
\end{equation}
whose extrema are obtained 
in the usual manner by computing the directional derivatives of $\mathcal{A}$ at $\Psi\in \mathfrak{X}$ in the directions of all fields $\eta$ that vanish on $\partial\Omega$.  We have that
\begin{eqnarray}\label{eq:32}
\lefteqn{\frac{d}{d\epsilon}\mathcal{A}(\Psi+\epsilon \eta)\Big|_{\epsilon=0}=
\int_\Omega [ \mathcal{L}_u(x,(\mathcal{N}\Psi)(x),\nabla(\mathcal{N}\Psi)(x))(\mathcal{N}\eta)(x)\hspace*{.5in}}\hspace{1.2in}\nonumber\\
&&+\mathcal{L}_v(x,(\mathcal{N}\Psi)(x),\nabla(\mathcal{N}\Psi)(x)) \nabla (\mathcal{N}\eta)(x)]  \,d^4x.
\end{eqnarray}
After integration by parts on the second summand and using the assumption that the boundary term vanishes, we find that $\Psi$ renders $\mathcal{A}$ stationary if the integral
\begin{eqnarray}\label{eq:33}
 \lefteqn{\int_\Omega  [\mathcal{L}_u(\cdot,(\mathcal{N}\Psi),\nabla(\mathcal{N}\Psi))(x)}\hspace{1in}
\nonumber\\
&& {}-\nabla (\mathcal{L}_v(\cdot,(\mathcal{N}\Psi),\nabla(\mathcal{N}\Psi))(x)] (\mathcal{N}\eta)(x)\,d^4x
\end{eqnarray}
vanishes for every $\eta$. 
Because $\mathcal{N}$ is invertible, it preserves the set of field variations  that vanish on $\partial\Omega$. Therefore, it follows from Eq.~\eqref{eq:33} that  \begin{equation}\label{eq:34} \mathcal{L}_u(\cdot,(\mathcal{N}\Psi),\nabla(\mathcal{N}\Psi))(x)-\nabla\mathcal{L}_v(\cdot,(\mathcal{N}\Psi),\nabla(\mathcal{N}\Psi))(x)=0.
\end{equation}
Eq.~\eqref{eq:34} is the nonlocal equation of motion; it is exactly the equation obtained by composition of the local Euler-Lagrange equation with the nonlocal operator $\mathcal{N}$.

If desired, the (Hilbert space) adjoint $\mathcal{N}^\dagger$ of the operator $\mathcal{N}$ may be  employed so that the integral~\eqref{eq:33} becomes
\begin{eqnarray}\label{eq:34.1}
\lefteqn{\int_\Omega  \mathcal{N}^\dagger[\mathcal{L}_u(\cdot,(\mathcal{N}\Psi),\nabla(\mathcal{N}\Psi))}\hspace{1in}\nonumber\\
&&{}-\nabla (\mathcal{L}_v(\cdot,(\mathcal{N}\Psi),\nabla(\mathcal{N}\Psi))](x) \eta(x)  \,d^4x
\end{eqnarray}
with the corresponding Euler-Lagrange equation
\begin{equation}\label{eq:35}
\mathcal{N}^\dagger[\mathcal{L}_u(\cdot,(\mathcal{N}\Psi),\nabla(\mathcal{N}\Psi))-\nabla (\mathcal{L}_v(\cdot,(\mathcal{N}\Psi),\nabla(\mathcal{N}\Psi))](x)=0.
\end{equation}
Because $\mathcal{N}$ is invertible, $\mathcal{N}^\dagger$ is also invertible. Thus,  Eqs.~\eqref{eq:34} and~\eqref{eq:35} are equivalent; that is, they have the same solutions. In general, Eq.~\eqref{eq:34} is simpler.   

For the massless scalar field, the integral in Eq.~\eqref{eq:33} is  
\begin{equation}\label{eq:36}
-\int_\Omega  \nabla\nabla(\mathcal{N}\Psi) (\mathcal{N}\eta)
\,d^4x=-\int_\Omega   \Box (\mathcal{N}\Psi) (\mathcal{N}\eta)
\,d^4x. 
\end{equation}
If the integral vanishes for all  $\eta$, then $\Box (\mathcal{N}\Psi) =0$, as previously demonstrated. 

The general approach presented here can be extended to the Dirac field;
this is discussed in  sections 5--8. We now turn to a derivation of the nonlocal
Maxwell equations from an action principle involving a nonlocal Lagrangian.

\section{Nonlocal Maxwell Lagrangian}
Consider the local inertial Lagrangian
\begin{equation}\label{eq:37}
\mathcal{L}_M=\frac{1}{16\pi} f_{\mu\nu}f^{\mu\nu}
  -\frac{1}{8\pi}f^{\mu\nu}(\partial_\mu a_\nu-\partial_\nu a_\mu)+ j_\mu a^\mu,
\end{equation}
where $f_{\mu\nu}$ and $a_\mu$ are regarded as independent fields and $j_\mu$ is the current associated with charged particles. It follows from the corresponding Euler-Lagrange equations that 
\begin{eqnarray}
\label{eq:38}  f_{\mu\nu}&=&\partial_\mu a_\nu-\partial_\nu a_\mu,\\
\label{eq:39}  \partial_\nu f^{\mu\nu}&=& 4\pi j^\mu. 
\end{eqnarray}
As is well known~\cite{L17}, the connection between the vector potential $a_\mu$ and the antisymmetric Faraday tensor $f_{\mu\nu}$ in Eq.~\eqref{eq:38} accounts for the first pair of Maxwell's equations, while the second pair is given by Eq.~\eqref{eq:39}. 

The transition to the nonlocal electrodynamics of accelerated systems occurs via
\begin{eqnarray}
\label{eq:40}a_\mu(x) &=&A_\mu(x)+\int K_\mu^{\;\; \nu}(x,y) A_\nu(y)\,d^4y,\\
\label{eq:41} f_{\mu\nu}(x)&=& F_{\mu\nu}(x)+\int K_{\mu\nu}^{\quad \rho\sigma}(x,y) F_{\rho\sigma}(y)\,d^4y  , 
\end{eqnarray}
where the kernels $K_{\mu\nu}$ and $ K_{\mu\nu\rho\sigma}$ are determined by the acceleration of the congruence of noninertial observers~\cite{L11}. In this case, the field equations are obtained from the substitution of Eqs.~\eqref{eq:40}--\eqref{eq:41} in  Eqs.~\eqref{eq:38}--\eqref{eq:39}. Writing the former equations as 
\begin{equation}\label{eq:42}
a_\mu=\mathbf{n}A_\mu, \qquad f_{\mu\nu}=\mathbf{N}F_{\mu\nu},
\end{equation}
we have the nonlocal field equations
\begin{eqnarray}
\label{eq:43}\mathbf{N}F_{\mu\nu} &=& \partial_\mu (\mathbf{n}A_\nu)-\partial_\nu (\mathbf{n}A_\mu) ,\\
\label{eq:44}\partial_\nu(\mathbf{N}F^{\mu\nu}) &=& 4\pi j^\mu.
\end{eqnarray}

It is now straightforward to show, using the method of the previous section, that Eqs.~\eqref{eq:43}--\eqref{eq:44} follow directly from the variational principle based on the nonlocal Maxwell Lagrangian
\begin{eqnarray}
\label{eq:45}\mathcal{L}_{N\hspace{-0.02in}M} &=&\frac{1}{16\pi}(\mathbf{N} F_{\mu\nu})(\mathbf{N}F^{\mu\nu})\nonumber\\
  &&{} -\frac{1}{8\pi}(\mathbf{N}F^{\mu\nu})[\partial_\mu(\mathbf{n} A_\nu)-\partial_\nu (\mathbf{n} A_\mu)]+ j_\mu (\mathbf{n}A^\mu). 
\end{eqnarray}
That is, instead of working with field variations $\delta F_{\mu\nu}$ and $\delta A_\mu$ as in section 2, it is far simpler to work with $\delta (\mathbf{N}F_{\mu\nu})=\mathbf{N}\delta F_{\mu\nu}$  and $\delta(\mathbf{n}A_\mu)=\mathbf{n}\delta A_\mu$ as demonstrated in general in section 3. These latter variations, just as the original field variations, vanish on the boundary hypersurface $\partial \Omega$ due to the circumstance that the kernels in Eqs.~\eqref{eq:40}--\eqref{eq:41} are zero except in the finite spacetime domain $\Omega$ in which acceleration takes place. 

In our background global inertial frame, we have employed Cartesian coordinates throughout; however, it should be possible, in principle, to work in any admissible (curvilinear) system of coordinates in Minkowski spacetime. For instance, the nonlocal field equations~\eqref{eq:43}--\eqref{eq:44} may be transformed to any other  coordinate system based on the invariance of the forms $ A_\mu dx^\mu$ and $F_{\mu\nu} dx^\mu\wedge dx^\nu$.

\section{Nonlocal Dirac equation}
According to the fundamental inertial observers that are at rest in the background global inertial frame $\mathcal{S}$ in Minkowski spacetime, the Dirac equation is given by 
\begin{equation}\label{eq1}
(i\gamma^\mu\partial_\mu-m) \psi(x)=0,
\end{equation}
where $\gamma^\mu$ are the constant Dirac matrices in the standard representation~\cite{8}. 

An accelerated observer in $\mathcal{S}$ is characterized by an orthonormal tetrad frame $\lambda^\mu_{\hspace{0.1in}(\alpha)}(\tau)$ that propagates along its worldline such that 
\begin{equation}\label{eq2}
\frac{d}{d\tau} \lambda^\mu_{\hspace{0.1in}(\alpha)}=
    \Phi_{(\alpha)}^{\hspace{0.1in}(\beta)}(\tau)\lambda^\mu_{\hspace{0.1in}(\beta)}.
\end{equation} 
Here, $\Phi_{(\alpha)(\beta)}(\tau)$ is the invariant antisymmetric acceleration tensor. The observer's local frame is defined by its temporal axis given by the timelike unit vector $\lambda^\mu_{{\hspace{0.1in}}(0)}= dx^\mu/d\tau$ and its spatial frame given by the three spacelike unit vectors $\lambda^\mu_{{\hspace{0.1in}}(i)}$, $i=1,2,3$. The translational acceleration of the observer is given by the ``electric'' part of $\Phi_{(\alpha)(\beta)}$, i.e. $\Phi_{(0)(i)}=g_i$, while the rotation frequency of the spatial frame with respect to a locally nonrotating (i.e. Fermi-Walker transported) frame is given by the ``magnetic'' part of $\Phi_{(\alpha)(\beta)}$, i.e. $\Phi_{(i)(j)}=-\epsilon_{ijk}\,\omega^k$. Thus the translational acceleration $g^i(\tau)$ and the rotation frequency $\omega^i(\tau)$ completely determine the noninertial character of the observer. To avoid unphysical situations, we assume that henceforth the observer's acceleration is turned on at some initial instant $\tau_i$ and then turned off at a later time $\tau_f$ along the observer's worldline. 

The Dirac spinor according to the momentarily comoving inertial observers along the path of the accelerated observer is given by $\hat\psi(\tau)=\Lambda(\tau)\psi(\tau)$, where
\begin{equation}\label{eq3}
 \Lambda(\tau)=e^{-\int_{\tau_i}^\tau \kappa (\tau')d\tau'}\Lambda(\tau_i)
\end{equation}
and $\kappa $ is given by 
\begin{equation}\label{eq4}
 \kappa (\tau)=\frac{i}{4}\Phi_{(\alpha)(\beta)}(\tau) \sigma^{\alpha\beta}
\end{equation}
with 
\begin{equation}\label{eq5}
 \sigma^{\alpha\beta}= \frac{i}{2}[\gamma^\alpha, \gamma^\beta].
\end{equation}
A detailed treatment is contained in~\cite{1}.

For the accelerated observer, however, the Dirac spinor is $\hat{\Psi}(\tau)$
given by Eq.~\eqref{eq:1}. The corresponding kernel can be expressed as
\begin{equation}\label{eq7}
\hat{k} (\tau,\tau')=\kappa (\tau'),
\end{equation}
where 
\begin{equation}\label{eq8}
\kappa (\tau')=-\frac{d\Lambda(\tau')}{d\tau'} \Lambda^{-1}(\tau').
\end{equation}
Thus $\kappa (\tau)$ coincides in this case with the invariant matrix given by Eq.~\eqref{eq4}.

It follows from Eqs.~\eqref{eq:1} and~\eqref{eq:3} that
\begin{equation}\label{eq10}
\Psi(\tau)=\psi(\tau)+\int_{\tau_i}^\tau k(\tau,\tau')\psi(\tau')\,d\tau',
\end{equation}
where
\begin{equation}\label{eq11}
k(\tau,\tau')=\Lambda^{-1}(\tau) \kappa (\tau') \Lambda(\tau').
\end{equation}

To extend these considerations to a whole class of accelerated observers that occupy a certain spacetime domain $\Omega$ in $\mathcal{S}$,  Eq.~\eqref{eq:6} can be written for spinors as 
\begin{equation}\label{eq13}
\psi(x)=\Psi(x)+\int_{\Omega}\mathcal{K}(x,y)\Psi(y)\,d^4y.
\end{equation}
In physically reasonable situations, the relationship between $\psi(x)$ and $\Psi(x)$ is unique within $\Omega$; that is, this nonlocal relationship is invertible, so that there is a unique kernel $\mathcal{R}$ such that
\begin{equation}\label{eq13.1}
\Psi(x) = \psi(x) + \int_\Omega \mathcal{R} (x,y) \psi(y)\, d^4 y .   
\end{equation}
To show explicitly how relations~\eqref{eq13} and~\eqref{eq13.1} come about, we work out two simple examples of the kernels $\mathcal{K}$ and $\mathcal{R}$ in the following section based on the results given in~\cite{1}.
The nonlocal Dirac equation for the class of accelerated observers then follows from Eqs.~\eqref{eq1} and~\eqref{eq13} and can be written as 
\begin{equation}\label{eq14}
(i\gamma^\mu\partial_\mu-m)[\Psi(x)+\int_{\Omega}\mathcal{K}(x,y)\Psi(y)\,d^4y]=0.
\end{equation}

We will show that this equation can be derived from a variational principle of stationary action involving certain bilinear scalar functionals of the form
\begin{equation}\label{eq15}
\langle \phi,\chi\rangle=\int_\Omega\overline{\phi} \chi\,d^4x,
\end{equation}
where $\phi$ and $\chi$ are spinors, and $\overline{\phi}$ is the adjoint of $\phi$ defined, using its Hermitian conjugate, $\phi^\dagger$, by $\overline{\phi}=\phi^\dagger\gamma^0$. Moreover, for an operator $\mathcal{O}$, the adjoint operator $\mathcal{O}^*$ is defined in this case by the relation
$\langle \mathcal{O}\phi,\chi\rangle=\langle \phi,\mathcal{O}^* \chi\rangle$ with respect to the nondegenerate inner product given by Eq.~\eqref{eq15}. The action for Eq.~\eqref{eq14} is introduced in section 8. We now turn to illustrative examples of kernels $\mathcal{K}$ and  $\mathcal{R}$.

\section{Examples}
Imagine first a class of rotating observers that are always at rest in $\mathcal{S}$.  These observers refer their measurements to the standard inertial axes in
$\mathcal{S}$ for $-\infty < t < t_i$, while for $t_i < t < t_f$ they ``rotate"
uniformly, since they refer their measurements to axes that rotate with
constant frequency $\omega$ about the $z$ direction. For the sake of simplicity, we
assume that these observers occupy the interior of a sphere of radius $R$ about the origin of spatial coordinates. The acceleration tensor is nonzero in this case only during the interval $(t_i,t_f)$, when the spatial frames undergo uniform rotation of frequency $\omega$ about the $z$ direction. We note that $\tau=t$, since each observer is fixed in space. This special class of rotating observers has been discussed in detail in section IV of~\cite{1}. We are interested in these observers until time $T\gg t_f$. The relevant kernels in Eqs.~\eqref{eq:1},~\eqref{eq10}, and~\eqref{eq:4} are given respectively by $\kappa=-(i\omega/2)\sigma^{12}$  and 
\begin{equation}\label{eq16}
k(t,t')=-\frac{i}{2}\omega\,\mathrm{diag}\,(q, -\frac{1}{q},q, -\frac{1}{q}),\quad r(t,t')=-\kappa,
\end{equation} 
where
\begin{equation}\label{eq17}
q=e^{i\omega (t'-t)/2}. 
\end{equation} 
It follows that $\mathcal{K}(x,x')=\mathcal{K}(t,\mathbf{x};t',\mathbf{x'})$, where
\begin{equation}\label{eq18}
\mathcal{K}(t,\mathbf{x};t',\mathbf{x'})=r(t, t')\,\mathcal{U}_{(t_i,T)}(t)\mathcal{U}_{(t_i,t_f)}(t') u(R-|\mathbf{x}|)\delta(\mathbf{x}'-\mathbf{x}).
\end{equation} 
Here $u(t)$ is the unit step function such that $u(t)=1$ for $t> 0$ and $u(t)=0$ for $t< 0$, and $\mathcal{U}_{(a,b)}$ is the unit bump function given by
\begin{equation}\label{eq19}
\mathcal{U}_{(a,b)}(t)=u(t-a)-u(t-b).
\end{equation}
Similarly, in this case 
\begin{equation}\label{eq21}
\mathcal{R}(t,\mathbf{x};t',\mathbf{x'})=k(t, t') \,\mathcal{U}_{(t_i,T)}(t)\mathcal{U}_{(t_i,t_f)}(t')u(R-|\mathbf{x}|)\delta(\mathbf{x}'-\mathbf{x}).
\end{equation}
Thus for these special rotating observers, $\Omega = (t_i, T) \times \Sigma_0$, where $\Sigma_0$ is the interior of the sphere of radius $R$.

Consider next observers $\mathcal{S}$ that are at rest in a finite region of space for $-\infty<t<t_i$ and at time $t_i$ start accelerating from rest along the $z$ direction with constant acceleration $g>0$. This acceleration is turned off at $t_f$ and the observers then move uniformly along the $z$ direction with speed $\beta_f=(t_f-t_i)/\zeta(t_f)$ until time $T\gg t_f$. Here the function $\zeta$ is given by 
\begin{equation}\label{eq21.1}
\zeta(t)=\sqrt{(t-t_i)^2+\frac{1}{g^2}}.
\end{equation}
These linearly accelerated observers have been discussed in section V of~\cite{1}.
Let $\Sigma$ denote the finite open region of space occupied by the trajectories of these observers for the interval of time $(t_i,T)$ that is of interest here.
Thus, in this case $\Omega = (t_i, T)\times\Sigma$. Using the results given in~\cite{1}, it is possible to show that $r=(i/2)g\sigma^{03}$ and 
\begin{equation}\label{eq21.2}
k=\frac{g}{2}[\sinh(\frac{\theta-\theta'}{2})-i \cosh(\frac{\theta-\theta'}{2})\sigma^{03}],
\end{equation}
where 
\begin{equation}\label{eq21.3}
\theta(t)=\ln\{g[t-t_i+\zeta(t)]\}
\end{equation}
and $\theta'=\theta(t')$. It follows from the arguments presented in~\cite{1} that the kernels $\mathcal{K}$ and $\mathcal{R}$ can then be written as
\begin{equation}\label{eq21.4}
\mathcal{K}(t,\mathbf{x}; t',\mathbf{x}')=r\frac{1}{g\zeta'} \mathcal{U}_{(t_i,T)}(t)\mathcal{U}_{(t_i,t_f)}(t')\mathcal{X}_\Sigma(\mathbf{x})\delta(x'-x)\delta(y'-y)\delta(z'-z+\zeta-\zeta')
\end{equation}
and
\begin{equation}\label{eq21.5}
\mathcal{R}(t,\mathbf{x}; t',\mathbf{x}')=k\frac{1}{g\zeta'} \mathcal{U}_{(t_i,T)}(t)\mathcal{U}_{(t_i,t_f)}(t')\mathcal{X}_\Sigma(\mathbf{x})\delta(x'-x)\delta(y'-y)\delta(z'-z+\zeta-\zeta'),
\end{equation}
where $\zeta'=\zeta(t')$ and $\mathcal{X}_\Sigma$ is the characteristic function of $\Sigma$. That is, $\mathcal{X}_\Sigma(\mathbf{x})$ is unity for $\mathbf{x}$ in $\Sigma$ and zero otherwise.  

\section{Local Dirac Lagrangian}
The Dirac equation can be obtained from the variational principle
\begin{equation}\label{eq22}
\delta\int \mathcal{L}_D\,d^4x=0,
\end{equation}
where the local Lagrangian density is given by 
\begin{equation}\label{eq23}
\mathcal{L}_D=\frac{1}{2}\overline{\psi} (i\gamma^\mu \partial _\mu-m)\psi-\frac{1}{2}[i(\partial_\mu\overline{\psi})\gamma^\mu+m\overline{\psi}]\psi.
\end{equation}
The field $\psi$ is complex, hence $\psi$ and $\overline{\psi}=\psi^\dagger\gamma^0$ are varied independently in Eq.~\eqref{eq22}. We note that the two parts of the Lagrangian~\eqref{eq23} are related by a total divergence; that is,
\begin{equation}\label{eq24}
\overline{\psi} (i\gamma^\mu \partial _\mu-m)\psi=i\partial_\mu J^\mu- [i(\partial_\mu\overline{\psi})\gamma^\mu+m\overline{\psi}]\psi,
\end{equation}
where $J^\mu=\overline{\psi}\gamma^\mu\psi$ is the current four-vector.
The principle of stationary action results in this case in the Dirac equation as well as the field equation for the adjoint spinor
\begin{equation}\label{eq25}
i(\partial_\mu\overline{\psi})\gamma^\mu+ m\overline{\psi}=0.
\end{equation}
Thus $\mathcal{L}_D$ vanishes when the Dirac equation is satisfied. It is straightforward to show that $\mathcal{L}_D$ is real and invariant under Lorentz transformations; in fact, these properties follow from the identity $\gamma^{\mu\dagger}\gamma^0=\gamma^0\gamma^\mu$ and the circumstance that under an inhomogeneous Lorentz transformation $x^\alpha=L^\alpha_{\hspace{0.1in}\beta}x^{'\beta}+s^\alpha $, we have
\begin{equation}\label{eq26}
\psi'(x')=S\psi(x), \qquad
\overline{\psi}'(x')=\overline{\psi}(x) S^{-1},
\end{equation}
where the spin transformation matrix $S$ depends on $L$,  $S\gamma^\alpha S^{-1}= L^\alpha_{\hspace{0.1in}\beta}\gamma^\beta$, and is such that $S^{-1}=\gamma^0S^\dagger \gamma^0$~\cite{8}. 

The energy-momentum tensor $T_{\mu\nu}$ associated with the Dirac spinor is given by
\begin{equation}\label{eq27}
T_{\mu}^{\hspace{0.1in}\nu}=-\mathcal{L}\, \eta_{\mu}^{\hspace{0.1in}\nu}+\frac{\partial\mathcal{L}}{\partial(\partial_\nu\psi)}\partial_\mu\psi+\partial_\mu\overline{\psi}\frac{\partial\mathcal{L}}{\partial(\partial_\nu\overline{\psi})}. 
\end{equation}
We note that $\mathrm{tr}\,(T_{\mu\nu})=m\overline{\psi} \psi$. Since $\mathcal{L}$ vanishes for the Dirac spinor, we find 
\begin{equation}\label{eq28}
T_{\mu\nu}=\frac{i}{2}[\overline{\psi}\gamma_\nu\partial_\mu\psi-(\partial_\mu \overline{\psi}) \gamma_\nu \psi],
\end{equation}
which is real and satisfies the conservation law $T_{\mu \hspace{0.1in},\, \nu}^{\hspace{0.1in}\nu}=0$. A symmetric energy-momentum tensor is given by
\begin{equation}\label{eq29}
{T}_{(\mu\nu)}=\frac{1}{2}( T_{\mu\nu}+T_{\nu\mu}),
\end{equation}
which is also real and conserved. The latter property follows because $\Box{\psi}=-m^2\psi$.

\section{Nonlocal Dirac Lagrangian}
To derive the nonlocal Dirac equation~\eqref{eq14} from a variational principle, we follow the general procedure outlined in section~3. Let us define a nonlocal operator $N$ such that $\psi=N\Psi$; that is,
\begin{equation}\label{eq30}
N\Psi(x)=\Psi(x)+\int_\Omega \mathcal{K}(x,y)\Psi(y) \,d^4y.
\end{equation}
On physical grounds, as described in section 1, $N$ is assumed to be invertible. That is, $\Psi=N^{-1}\psi$, where 
\begin{equation}\label{eq31.1}
N^{-1}\psi(x) = \psi(x) + \int_\Omega \mathcal{R} (x,y) \psi(y)\, d^4 y .   
\end{equation}
Moreover, under an inhomogeneous Lorentz transformation $(x \mapsto x', y \mapsto y')$, we have in analogy with Eq.~\eqref{eq26},
\begin{equation}\label{eq31.2}
\Psi'(x')=S \Psi,\qquad \overline{\Psi}'(x')=\overline{\Psi}(x) S^{-1}   
\end{equation} 
and therefore
\begin{equation}\label{eq31.3}
\mathcal{K}'(x',y')=S \mathcal{K}(x,y)S^{-1},\qquad 
\mathcal{R}'(x',y')=S \mathcal{R}(x,y)S^{-1}.   
\end{equation}

The desired Lagrangian is obtained from the substitution of $\psi$ with $N\Psi$ in the local Dirac Lagrangian $\mathcal{L}_D$, namely,
\begin{equation}\label{eq31}
\mathcal{L}_{N\hspace{-0.02in}D}=\frac{1}{2} \overline{N\Psi}(i\gamma^\mu\partial_\mu-m) N\Psi-\frac{1}{2} [i(\partial_\mu \overline {N\Psi})\gamma^\mu+m \overline{N\Psi}]       N\Psi.
\end{equation}
Here $\overline{N\Psi}$ is the usual adjoint spinor obtained from Eq.~\eqref{eq30}, i.e.
\begin{equation}\label{eq32}
\overline{N\Psi}(x)=\overline{\Psi}(x)+\int_\Omega \overline{\Psi}(y) \overline{\mathcal{K}}(x,y)\,d^4y,
\end{equation}
where
\begin{equation}\label{eq33}
 \overline{\mathcal{K}}(x,y)=\gamma^0 \mathcal{K}^\dagger(x,y)\gamma^0.
\end{equation}
On the other hand, the adjoint of $N$ is given by
\begin{equation}\label{eq34}
{N^* \Psi}(x)={\Psi}(x)+\int_\Omega\overline{\mathcal{K}}(y,x) {\Psi}(y)\,d^4y,
\end{equation}
as can be easily checked using $\langle N\phi,\chi\rangle=\langle \phi, N^*\chi\rangle$. We note that $N^*$ is invertible because $N$ is invertible.

It is now straightforward to show, using the formalism in section~3, that the variational principle  $\delta \mathcal{A}=0$ with $\mathcal{A}= \int_\Omega \mathcal{L}_{N\hspace{-0.02in}D}\,d^4x$ leads to the nonlocal Dirac equation~\eqref{eq14}. To see how this comes about, we first vary $\overline{\Psi}$ while $\Psi$ is kept fixed; then, 
\begin{equation}\label{eq36a}
\delta\mathcal{A}=\int_\Omega \delta\overline{\Psi}(x)V(x)\,d^4x-\frac{i}{2}\int_\Omega \frac{\partial}{\partial x^\mu}[\delta\overline{\Psi}(x)\gamma^\mu N\Psi(x)]\,d^4x,   
\end{equation}
where
\begin{equation}\label{eq37a}
V(x)=N^*(i\gamma^\mu \partial_{\mu}-m) N\Psi-\frac{i}{2}\int_\Omega \frac{\partial}{\partial y^\mu}[\overline{\mathcal{K}}(y,x) \gamma^\mu N\Psi(y)]\,d^4y.   
\end{equation}
The second integral in~\eqref{eq36a} vanishes via Stokes' theorem, since we assume that $\delta\overline{\Psi}=0$ on $\partial \Omega$. Similarly, the second term in Eq.~\eqref{eq37a} is zero due to the fact that by construction $\mathcal{K}(x,y)$ vanishes for $x\in \partial\Omega$. Thus, in this case 
$\delta\mathcal{A}=0$ implies that $V=0$. The operator $N^*$ is invertible, hence we recover the nonlocal Dirac equation~\eqref{eq14}. Let us now vary $\Psi$ while $\overline{\Psi}$ is fixed; then, 
\begin{equation}\label{eq38a}
\delta\mathcal{A}=-\int_\Omega W(x) \delta \Psi(x) \,d^4x+\frac{i}{2}\int_\Omega \frac{\partial}{\partial x^\mu}[\overline{N\Psi}(x)\gamma^\mu \delta \Psi(x)]\,d^4x,   
\end{equation}
where
\begin{equation}\label{eq39a}
W(x)=Z(x)+\int_\Omega Z(y) \mathcal{K}(y,x) \,d^4y-\frac{i}{2}\int_\Omega \frac{\partial}{\partial y^\mu}[\overline{N\Psi}(y)\gamma^\mu \mathcal{K}(y,x)]\,d^4y   
\end{equation}
and
\begin{equation}\label{eq40a}
Z=i(\partial_\mu \overline{N\Psi})\gamma^\mu +m \overline{N\Psi}.  
\end{equation}
Here we assume that $\delta\Psi=0$ on $\partial\Omega$. As before, the last integrals in Eqs.~\eqref{eq38a} and~\eqref{eq39a} vanish and we find that $\delta \mathcal{A}=0$ implies that $W=0$, i.e.
\begin{equation}\label{eq41a}
Z(x)+ \int_\Omega  Z(y)\mathcal{K}(y,x) \,d^4y=0.   
\end{equation}
We note that the adjoint of Eq.~\eqref{eq34} can be written as 
\begin{equation}\label{eq42a}
\overline{N^*\Psi}(x)=\overline{\Psi}(x)+\int_\Omega\overline{\Psi}(y) \mathcal{K}(y,x) \,d^4y.  
\end{equation}
Therefore, 
\begin{equation}\label{eq43a}
\overline{N^*\gamma^0 Z^\dagger}(x)=Z(x)+\int_\Omega Z(y) \mathcal{K}(y,x) \,d^4y.  
\end{equation}
Hence Eq.~\eqref{eq41a} implies that $Z=0$, since $N^*$ and $\gamma^0$ are both invertible. Thus we find the nonlocal Dirac equation for $\overline{\Psi}$.

These considerations suggest that a natural way to define a conserved nonlocal energy-momentum tensor $\mathcal{T}_{\mu\nu}$ for accelerated systems is via the substitution of $\psi$ with $N\Psi$ in the local tensor $T_{\mu\nu}$. The result is 
\begin{equation}\label{eq35}
\mathcal{T}_{\mu\nu}=\frac{i}{2}[\overline{N\Psi}\gamma_\nu\partial_\mu(N\Psi)-(\partial_\mu\overline{N\Psi})\gamma_\nu N\Psi],
\end{equation}
which is real and conserved; that is $\mathcal{T}^{\;\nu}_{\mu\hspace{0.05in} ,\nu}=0$. As before, $\mathcal{T}_{(\mu\nu)}$ provides a \emph{symmetric}, real, and conserved energy-momentum tensor for the Dirac field. 

The currents of energy and momentum measured by an accelerated observer are obtained from the projection of the energy-momentum tensor on the observer's tetrad frame; therefore, it follows from $\psi=N\Psi$ that the accelerated observer, at each instant along its worldline, measures the same currents of energy and momentum as the locally comoving observer. Thus in this sense, the notions of energy current and momentum current are locally defined, though their connection to the field is nonlocal for accelerated observers. 

\section{Discussion}
Lagrangian dynamics in discussed in this paper within the framework of the nonlocal theory of accelerated systems. Specifically, we show that in the case of acceleration-induced
nonlocality, the variational embedding problem~\cite{15} has a natural solution involving the composition of the local Lagrangian with the nonlocal transformation rule for the field.  Moreover, the basic idea developed in this work leads to a nonlocal measure of energy via the energy-momentum tensor that can be derived from the local inertial Lagrangian in the standard manner. Briefly, the local energy-momentum tensor composed with the nonlocal transformation rule for the field leads to a new tensor that when projected on the accelerated observer's tetrad frame would amount to the measured energy-momentum tensor of the field according to the noninertial observer.  

We have also solved the variational embedding problem~\cite{15} involving the nonlocal Dirac equation for accelerated systems in Minkowski spacetime. We have shown that this equation can be derived from an appropriate variational principle of stationary action. In the absence of acceleration, we recover the standard results of the Dirac theory for inertial observers.

It is possible to provide a formal solution to the variational embedding
problem for the nonlocal Dirac equation by using the method described in
chapter 2 of~\cite{15}. However, it is then necessary to introduce in this case
an auxiliary spinor. The general approach adopted in the
present work is more satisfactory from a physical point of view. Moreover,
it provides a natural way of introducing the energy-momentum tensor into the
nonlocal theory of accelerated systems.

 Finally, it is important to remark that the nonlocal theory of accelerated systems has thus far dealt with each nonlocal field equation separately. However, the main procedure developed here can be employed to study \emph{interacting} fields from the standpoint of accelerated systems.

\section*{Acknowledgments}
The work of C. Chicone  is supported in part by the grant NSF/DMS-0604331. B. Mashhoon is grateful to Friedrich Hehl and Dominic Edelen for valuable
discussions and correspondence, respectively.

\end{document}